\documentclass[pra,aps,amsmath,twocolumn,showpacs,floatfix]{revtex4-1}
\usepackage{graphicx}
\usepackage{bm}
\usepackage{pdfpages}
\input{epsf}
\begin{document}
\epsfverbosetrue
\def\la{{\langle}}
\def\ra{{\rangle}}
\def\vep{{\varepsilon}}
\newcommand{\beq}{\begin{equation}}
\newcommand{\eeq}{\end{equation}}
\newcommand{\beqa}{\begin{eqnarray}}
\newcommand{\eeqa}{\end{eqnarray}}
\newcommand{\q}{\quad}
\newcommand{\h}{\hat{H}}
\newcommand{\ha}{\hat{h}}
\newcommand{\p}{\partial}
\newcommand{\A}{|\Omega'|}
\newcommand{\AC}{{\it AC}}
\newcommand{\n}{\\ \nonumber}
\newcommand{\om}{\omega}
\newcommand{\Om}{\Omega}
\newcommand{\os}[1]{#1_{\hbox{\scriptsize {osc}}}}
\newcommand{\cn}[1]{#1_{\hbox{\scriptsize{con}}}}
\newcommand{\sy}[1]{#1_{\hbox{\scriptsize{sys}}}}

\title{Statistics of resonance and non-resonance tunnelling of fermionised cold atoms}
\author {D. Sokolovski$^{a,b}$}
\author {L.M. Baskin$^c$}
\affiliation{$^a$ Departmento de Qu\'imica-F\'isica, Universidad del Pa\' is Vasco, UPV/EHU, Leioa, Spain}
\affiliation{$^b$ IKERBASQUE, Basque Foundation for Science, E-48011 Bilbao, Spain}
\affiliation{$^c$The Bonch-Bruevich State University of Telecommunications,
 193232, Pr. Bolshevikov 22-1, Saint-Petersburg, Russia}
\begin{abstract}
We show that a short-range strong repulsive (contact) interaction between the particles in the barrier may change the statistics of two-particle tunnelling. In the case of a resonance of a width $\Gamma$, the effect would be observed if the time between the two impacts is of order of $\hbar/\Gamma$.
The statistics of non-resonance tunnelling across a broad potential barrier remain unaffected, 
which suggests that there is no appreciable delay in the classically forbidden region.
\end{abstract}
\maketitle
\vskip0.5cm

\section{Introduction}
\noindent
Over the years there has been a considerable interest in transport of photons and electrons in optical waveguides and quantum wires \cite{WG}. Varying the diameter of a wire one can achieve resonance and non-resonance tunnelling conditions. For carriers with energies between the first and the second thresholds such tunnelling is a quasi one-dimensional process. Similar quasi one-dimensional confinement  can be constructed for cold atoms with the help of laser based techniques \cite{May}.

Recently, highly accurate calculations have been reported for single-particle tunnelling in such structures \cite{BASK}. However, equally accurate evaluation of the effects of interaction between the carriers remains an open problem. Such effects (e.g., the Coulomb blockade for tunnelling electrons \cite{CB1}-\cite{CB3}) are often described in terms of the tunnelling (transfer) Hamiltonian (TH) \cite{TH}. However, no rigorous scheme for constructing a TH from one-particle scattering states is known to date \cite{TH2}.

The effects of collisions between transmitted particles can be evaluated exactly in the case of a contact potential describing a very short range strong repulsive interaction between bosonic atoms in Tonks-Gerardeu (TG) gas \cite{TG}. There, as in the case of non-interacting identical particles, the final state of the system can be obtained by simple symmetrisation once all single-particle evolutions are known \cite{Feyn},  \cite{Klich}. Several authors have studied the escape of TG atoms from a potential trap, \cite{Muga} - 
\cite{Germans2}. In this Brief Report we study how inter-particle interaction, it its most elementary form, may influence scattering of such atoms. 
We also analyse the conditions under which  two TG atoms incident on a barrier from the same side may meet in the barrier region.
 A classical analogy would be that of the second particle coming within the interaction range, as the first particle is slowed down, e.g., passing over a potential hill.
A related question about the amount time for which a tunnelled particle is detained in the barrier has been under discussion since early 1930's, and we refer the reader to several reviews of the subject \cite{REV}-\cite{REV2}.
\section {More about the problem }
\noindent
We start with a brief discussion of what may happen if two particles with sufficiently low energies  are sent towards a potential barrier. 
Figure 1 sketches a scenario for two identical impenetrable classical particles of mass $m$ with the coordinates $x_1$ and $x_2$.
The first particle arrives at the barrier, is reflected, and on its way back collides with the second one. 
The two particles exchange momenta, the first particles returns to the barrier and is reflected for the second time. After that both particles leave, and their joint position is labelled (1) in the $(x_1,x_2)$-plane.
\newline
Quantally, one may try to address the same problem by constructing the scattering eigenstates of the Hamiltonian (we use $\hbar=1$)
\begin{eqnarray}\label{3}
\hat{H}=-\partial^2_{x_1}/2m-\partial^2_{x_1}/2m 
+V(x_1)+V(x_2)+U(|x_1-x_2|), \q\q
\end{eqnarray}
where $V$ is the barrier potential, and $U$ is a short-range interaction between the particles.
\newline
For our purpose it is, however, more convenient to consider two distinguishable particles represented by wave  packets. 
Since wave packets may also tunnel, we must add three more scenarios, also shown in Fig.1
After scattering is completed, 
the wave function in the $(x_1,x_2)$-plane evolves into four non-overlapping parts, corresponding to the four exclusive outcomes in Fig.1.
\newline 
The diagram in Fig.1 is only schematic, as we do not know whether a transmitted particle spends in the barrier an appreciable amount of time. We can try to test it in the following way. If the initial wave packets are well separated in time and space, the two particles should tunnel sequentially, the probability of the double transmission process unaffected by the 
interaction $U$. If, on the other hand, the particle 1 is detained in the barrier region, the particle 2 can catch up with it there. Then the inter-particle collisions may change the probabilities for the four outcomes shown in Fig.1. We will study the conditions under which this can happen.
\newline
\begin{figure}
	\centering
		\includegraphics[width=4cm,height=4cm]{{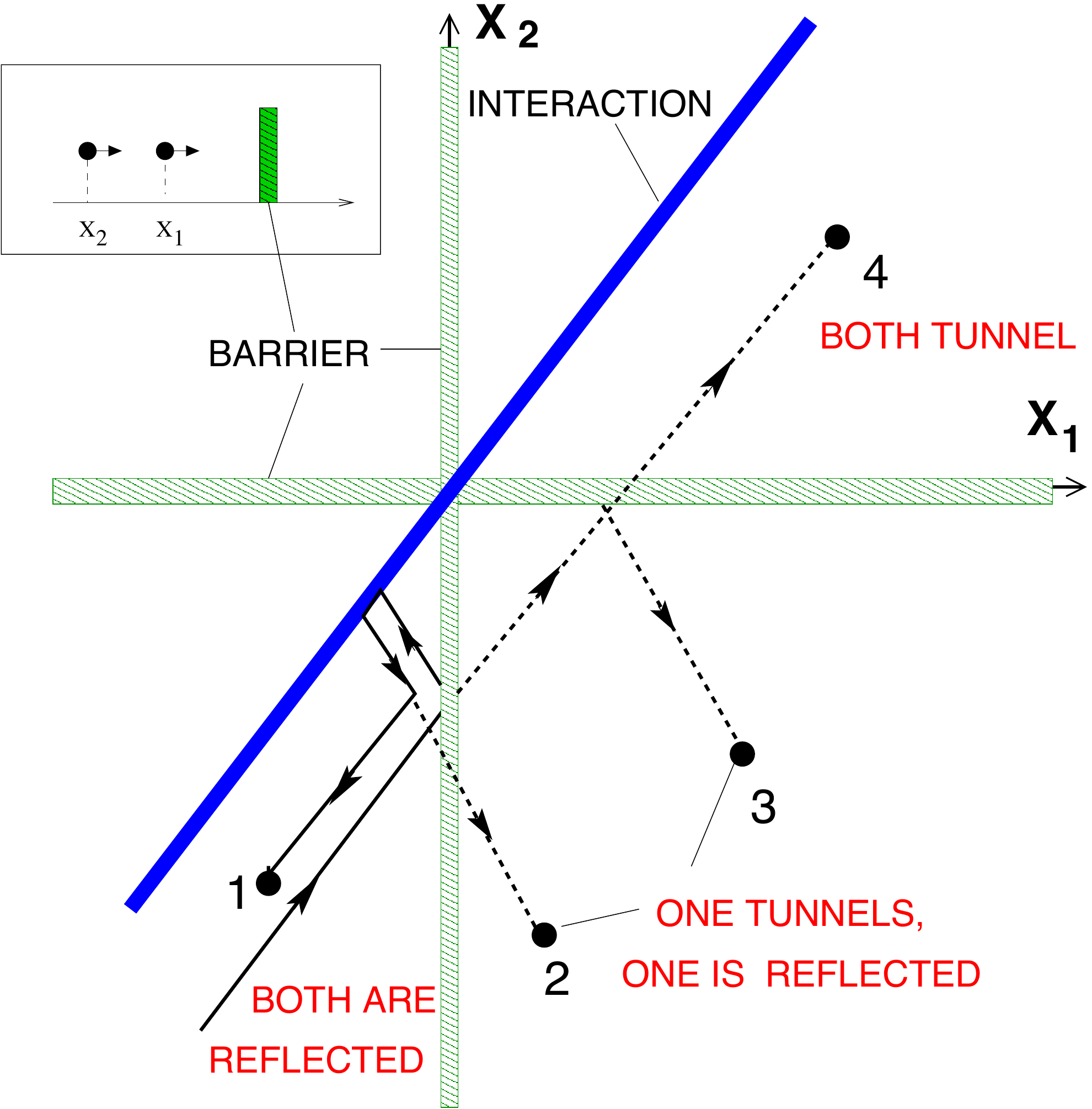}}
\caption{ Schematic diagram in the ($x_1,x_2$)-plane illustrating possible outcomes if  two distinguishable impenetrable particles 
interacting via a short range potential (thick diagonal line) are  scattered by a 
potential barrier (shaded): (1) both particle are reflected; (2,3) the first particle tunnels, and the second is 
reflected, and (4) both particles tunnel.
}
\label{fig:4}
\end{figure}

\section {Fermionised bosons and identical particles. }
\noindent
The two-particle tunnelling problem is readily solved for identical bosonic atoms interacting via the TG contact potential
\begin{eqnarray}\label{2}
U^{TG}(|x_1-x_2|)=\lim_{\gamma\to \infty}\gamma\delta(x_1-x_2). 
\end{eqnarray}
The potential (\ref{2}) forbids two particles to occupy the same location in space, and its effects are broadly similar to those of the Pauli principle for fermions. The wave function for two such 'fermionised bosons' is just the antisymmetrised combination of one-particle states,  \cite{TG},
\begin{eqnarray}\label{12}
\Psi^{FB}(x_1,x_2,t)=N^{-1}s(x_1,x_2)\times,\q\q\q\q \n
[\Psi_1(x_1,t)\Psi_2(x_2,t)-\Psi_1(x_2,t)\Psi_2(x_1,t)],\q
\end{eqnarray}
where the factor $s(x_1,x_2)=1$ for $x_1>x_2$, and $-1$ otherwise, serves to give $\Psi^{FB}(x_1,x_2,t)$ overall bosonic symmetry, $\Psi^{FB}(x_1,x_2,t)=\Psi^{FB}(x_2,x_1,t)$, $N$ is the normalisation.
In Eq.(\ref{12}), $\Psi_1(x_1,t)$ and $\Psi_2(x_2,t)$ are the one-particle states obtained by the evolution of the initial wave packets with the Hamiltonian (\ref{3}) where $U$ is put to zero, $U\equiv 0$. 
Thus,
to know the state of two TG atoms after scattering, it is sufficient to solve first two one-particle tunnelling problems, and apply the symmetrisation procedure  (\ref{12}) in the end.
\newline
To evaluate the likelihoods of the tunnelling outcomes,
 we will require the probability to find the fermionised atoms in the region $a\le x_1, x_2\le b$, $P^{FB}_{ab}(t)=\int_a^b dx_1\int_c^d dx_2|\Psi^{FB}(x_1,x_2,t)|^2$. From Eq.(\ref{12}) we have
\begin{eqnarray}\label{13}
P^{FB}_{}(t) \equiv P_0 - \delta P=\q\q\q\q\q\q\q\q\q\n
2N^{-2}[I_{}(\Psi_1,\Psi_1,t)I_{}(\Psi_2,\Psi_2,t)-|I_{}(\Psi_1,\Psi_2,t)|^2]\n
\end{eqnarray}
where
\begin{eqnarray}\label{14}
I_{}(\Psi_i,\Psi_j,t)\equiv \int_a^bdx\Psi^*_i(x,t)\Psi_j(x,t), \q i,j=1,2.\q
\end{eqnarray}
The first term in the square brackets is just the joint probability for two independent particles.
It is reduced by the exchange term, which results from contact interaction (\ref{2}).
By construction, Eqs.(\ref{13})-(\ref{14}) apply also to scattering of neutral fermions (spinless, or prepared in the same spin state). They are also valid for non-interacting bosons, provided the sign of the exchange term in (\ref{13}) is changed to a 'plus'.
\section {Two-particle tunnelling}
\noindent
Next we assume that a source at $x=0$ emits at $t=0$ an atom in a Gaussian state of a coordinate width $\sigma$ with a mean momentum $p_0$ towards a finite-width potential barrier $V(x)$ located further to the right. After a time $T$ an identical atom  is launched in exactly the same state. The atoms interact via the TG potential (\ref{2}), and their energy is below the barrier top. 
Using the results of the previous Section, we will study the dependence of the probabilities for the outcomes shown in Fig.1 on the time $T$ between the emissions.
\newline
Before the atoms arrive at the barrier, the two one-particle states are given by,
\begin{eqnarray}\label{15}
\nonumber
\Psi_1(x_1,t)= (2\pi)^{-1/2}\int dp A(p) \exp[ipx-ip^2(t+T)/2m],
\\
\Psi_2(x_2,t)=(2\pi)^{-1/2}\int dp A(p) \exp[ipx_2-ip^2t/2m],
\q\q\vspace {1mm}
\end{eqnarray}
where $A(p)$ is a Gaussian peaked at $p=p_0$, 
\begin{eqnarray}\label{15a}
A(p)=\sigma^{1/2}/(2\pi)^{1/4}\exp[-(p-p_0)^2\sigma^2/4], 
\end{eqnarray} and the spread of the momenta around $p_0$ is of order of $1/\sigma$.
\newline
There should be no interaction between the atoms at launch. 
If so, the exchange term in Eq.(\ref{12}) must vanish 
\begin{eqnarray}\label{16}
\delta P^{init}(T)=
|\int dp |A(p)|^2 \exp(-ip^2T/2m)|^2
\approx 0,\q\
\end{eqnarray}
and for the normalisation constant in (\ref{12}) we have $N=\sqrt{2}$.
The r.h.s in Eq.(\ref{16}) vanishes for large $T$'s because the oscillations of the exponential 
become rapid compared to $1/\sigma$.
Neglecting spreading of the wave packets over the time $T$, we write
$\exp(-ip^2T/2m)\approx  \exp-ip_0^2T/2m-ip_0T(p-p_0)/m]$, in which case Eq.(\ref{16}) requires that the initial  distance between the atoms, $p_0T/m$, must exceed their wave packet width $\sigma$.
Equation (\ref{16}) will then hold until the first atom reaches the barrier.
\newline
After both atoms have left the barrier region, each wave packet splits into the (non-overlapping) reflected (R) and transmitted (T) parts, 
\begin{eqnarray}\label{17}
\Psi_i(x_i,t)=\Psi^T_i(x_i,T)+\Psi^R_i(x_i,T),\q i=1,2,\q
\end{eqnarray} obtained by multiplying each plane wave in (\ref{15}) by the transmission, $B^T(p)$, or the reflection, $B^R(p)$, amplitudes, $A(p)\to A(p)B^{T,R}(p)$.
There are three probabilities for three possible outcomes, $P_{TT}$ (both atoms tunnel), $P_{RT}$ (one atoms tunnels and one is reflected), and $P_{RR}$ (both atoms are reflected).
Evaluating the integrals over the corresponding quadrants of the $(x_1,x_2)$-plane with the help of (\ref{13}), (\ref{16}) and (\ref{17}) we find
\begin{eqnarray}\label{19}
P_{TT}=(P^T)^2\mp\delta P,\q\q\q\q\q\q\q\q\q\q\q\q\n
P_{RT}=2[P^TP^R \pm \delta P], \hspace{2 mm}\q\q\q\q\q\q\q\q\q\q\n
P_{RR}=(P^R)^2\mp\delta P,\q\q\q\q\q\q\q\q\q\q\q\q
\end{eqnarray}
where the lower sign corresponds to non-interacting bosons, $P^T$ and $P^R$ are the single-particle tunnelling and reflection probabilities,
$P^{R,T}(x_1,t)=
\int |A(p)|^2 |B^{R,T}(p)|^2dp,$
and
\begin{eqnarray}\label{20}
\delta P(T)= \huge{|}\int dp |B^T(p)|^2 |A(p)|^2 \exp(-ip^2T/2m)|^2.\q\q
\end{eqnarray}
Note that once both atoms have left the barrier region, $\delta P$ in Eq.(\ref{20}) 
is independent of the
time $t$. Thus, it must result from the interaction between the atoms in the barrier region.
\newline
Equations (\ref{19})  show that the short range repulsion (\ref{2}) can modify the statistics of two-particle tunnelling, with two TG atoms less likely to exit the barrier on the same side.
The mean number of tunnelled particles remains, however, unchanged 
$\la n \ra_T \equiv 2\times  P_{TT} + 1\times P_{TR}=2P^T$.
\section{Resonance tunnelling}
\noindent
With Eq.(\ref{16}) holding, the term $\delta P(T)$ would not vanish provided the momentum distribution of the tunnelled atom, $|B^T(p)A(p)|^2$, is much narrower than its initial one, $|A(p)|^2$. 
Consider, therefore, a barrier supporting several narrow resonances of the widths $\Gamma_n$ at the energies $E^r_n$, $n=1,2...$. Neglecting the background, 
we may write $|B^T(p)|^2$ in the Breit-Wigner form
\begin{eqnarray}\label{24}
|B^T(p)|^2=\sum_n\frac{\Gamma_n^2}{(p^2/2m-E^r_n)^2+\Gamma_n^2}.
\end{eqnarray}
An example is given in Fig.2, for a double-delta barrier \cite{DSC2}, $V(x)=\Om [\delta(x)+\delta(x-d)]$, where the dots show the approximation (\ref{24}) for the first two resonance peaks, $n=0,1$.
\begin{figure}
	\centering
		\includegraphics[width=6cm,height=4cm]{{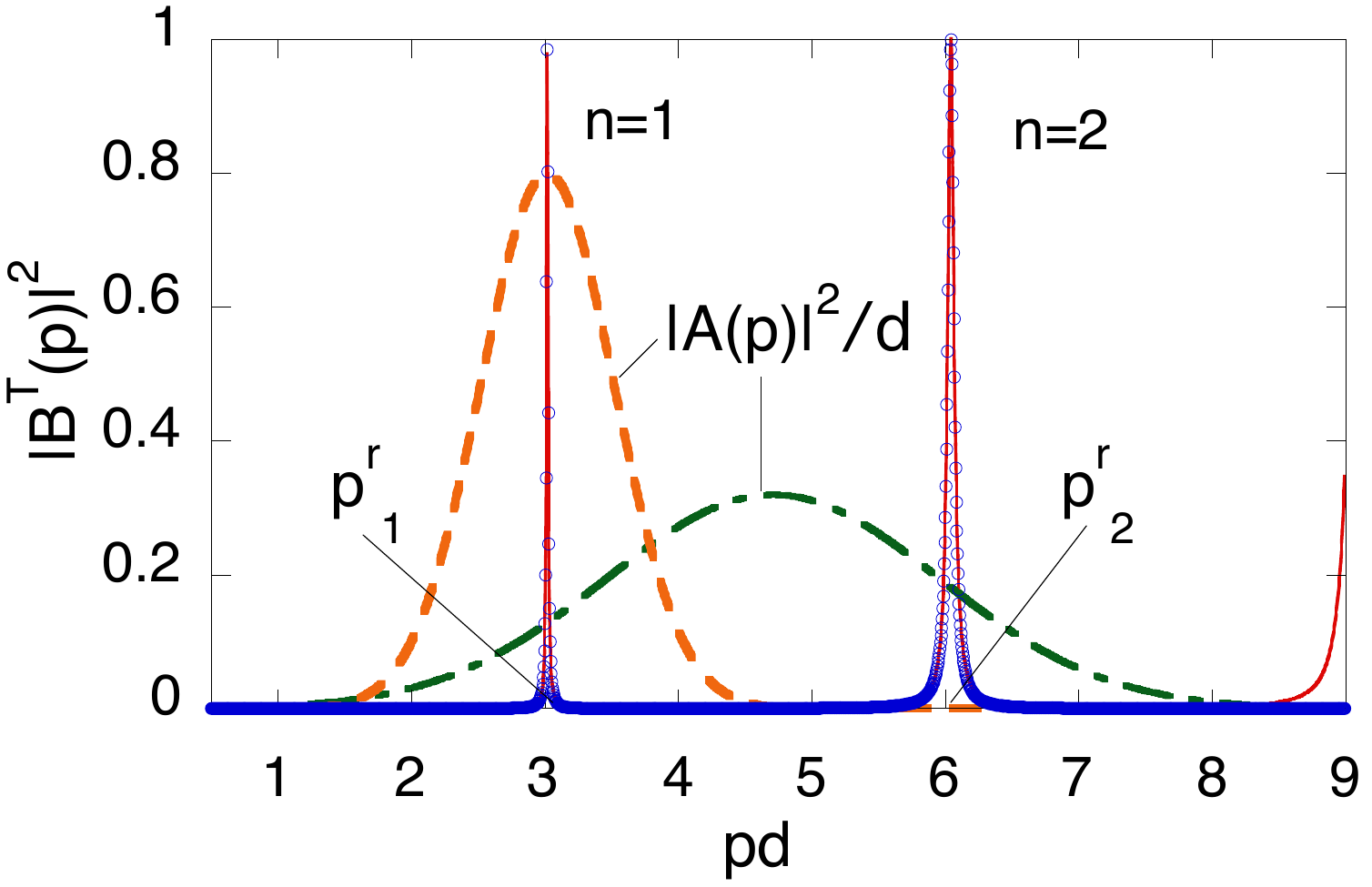}}
\caption{ Transparency of a double-delta barrier with $\Omega d=50$, (solid) and its Breit-Wigner approximation (\ref{24}) (circles). Also shown are the momentum distributions $|A(p)|^2$ which probe one (dashed) and two (dot-dashed) resonances.
}
\label{fig:4}
\end{figure}
\newline
Choosing the wave packet broad enough to probe just one resonance at $E^r_1$ , yet sufficiently narrow for its momentum distribution to be broader than the resonance peak (see Fig.2, dashed),  
we find
\begin{eqnarray}\label{25}
P_{TT}\approx (P^T)^2[1-\exp(-2\Gamma_1 T)], 
\end{eqnarray}
where 
$P^T=\frac{m\pi} {p^r_1}|A(p^r_1)|^2\Gamma_1$. 
The exchange term vanishes exponentially with $T$ (see Fig.3a), and the second atom catches up with the first in the barrier region provided $T\lesssim1/\Gamma_1$.
\newline
For a wave packet probing more than one resonance, there is possibility of interference effects. For example, if two resonances contribute to the one-particle transmission (see Fig.2, dot-dashed),
$\delta P(T)$ in Eqs.(\ref{19}) takes the form
\begin{eqnarray}\label{27}
\delta P(T) \approx \sum_{n=1,2}\frac{m^2\pi^2}{(p^r_n)^2}|A(p^r_n)|^4\Gamma_1^2 \exp(-2\Gamma_nT)+\n
2\frac{m^2\pi^2}{(p^r_1p^r_2)^2}|A(p^r_1)|^2|A(p^r_2)|^2\times\q\q\q\q\q
\n
 \exp[-(\Gamma_1+\Gamma_2)T]
\cos[(E_2^r-E_1^r)T].
\end{eqnarray}
As shown in Fig. 3b, it undergoes sinusoidal oscillations, 
so that the two-atom transmission probability is reduced further
whenever the time between the emissions, $T$, equals $2k\pi/(E_2^r-E_1^r))$, $k=1,2,...$.
\section{Non-resonance tunnelling}
\noindent
Finally, we consider two atoms with a mean momentum $p_0$ tunnelling across a rectangular barrier of a  height $V> p_0^2/2m$ and a width $d$ (the results are easily generalised to a barrier of an arbitrary form in the semi-classical approximation). For a broad barrier we have
\begin{figure}
	\centering
		\includegraphics[width=5cm,height=8cm]{{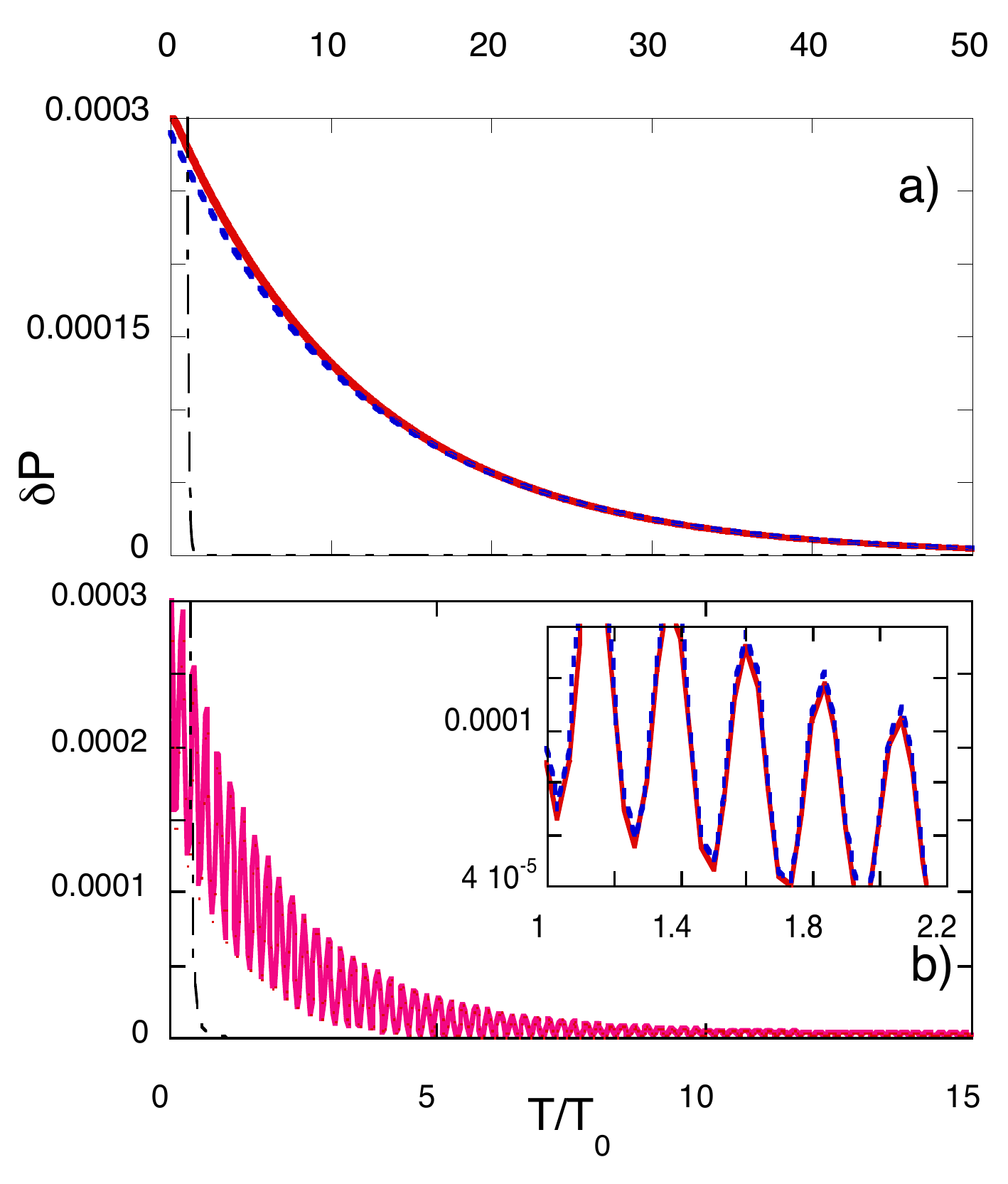}}
\caption{ $\delta P$ in Eq.(\ref{20}) (solid) for the double-delta potential barrier in Fig.2: (a) with just the $n=1$ resonance contributing (cf. Fig.2, dashed); and (b) with 
both the $n=1$ and $n=2$ resonances contributing (cf. Fig.2, dashed). The oscillations are magnified in the inset. Also shown are Breit-Wigner approximations 
(\ref{25}) and (\ref{27}) (dashed), and $\delta P^{init}(T)$ in Eq.(\ref{16}) (dot-dashed). 
 }
\label{fig:4}
\end{figure}
\begin{eqnarray}\label{27}
|T(p)|^2\approx C_1(p) \exp(-2S(p,d)),\q S(p,d)\equiv d\sqrt{2mV-p^2} \q\q
\end{eqnarray}
Expanding $S(p,d)$ to the second order in $(p-p_0)$ yelds
\begin{eqnarray}\label{28}
S(p,d)\approx S(p_0,d)+S'(p_0,d)(p-p_0)+S''(p_0,d)(p-p_0)^2/2. \q\q
\end{eqnarray}
Since $S''(p_0,d)=2mV/(2mV-p_0^2)^{1/2}$ is positive, there is no narrowing of the transmitted momentum distribution (see Fig.4a).
Neglecting, as before, the spreading of the wave packet, 
and evaluating the Gaussian integral in Eq.(\ref{20}) we have
\begin{eqnarray}\label{29}
P_{TT}\approx (P^T)^2\left \{1-\exp \left [ -\frac{T^2}{T_0^2}\frac{k_0^3\sigma^2}{\sigma^2k_0^3-2(p_0^2+k_0^2)d}\right ]\right \}, \q\q
\end{eqnarray}
where $k_0\equiv \sqrt{2mV-p_0^2}$ and $T_0\equiv m \sigma/p_0$.
For two atoms launched independently, we already have $T/T_0>>1$ (cf. Sect. IV). Since the second factor in the exponent is typically greater than unity, we have $P_{TT}\approx (P^T)^2$. Thus, two atoms, well separated initially, behave as if they never collide in a classically forbidden region.

\begin{figure}
	\centering
		\includegraphics[width=5.5cm,height=8.5cm]{{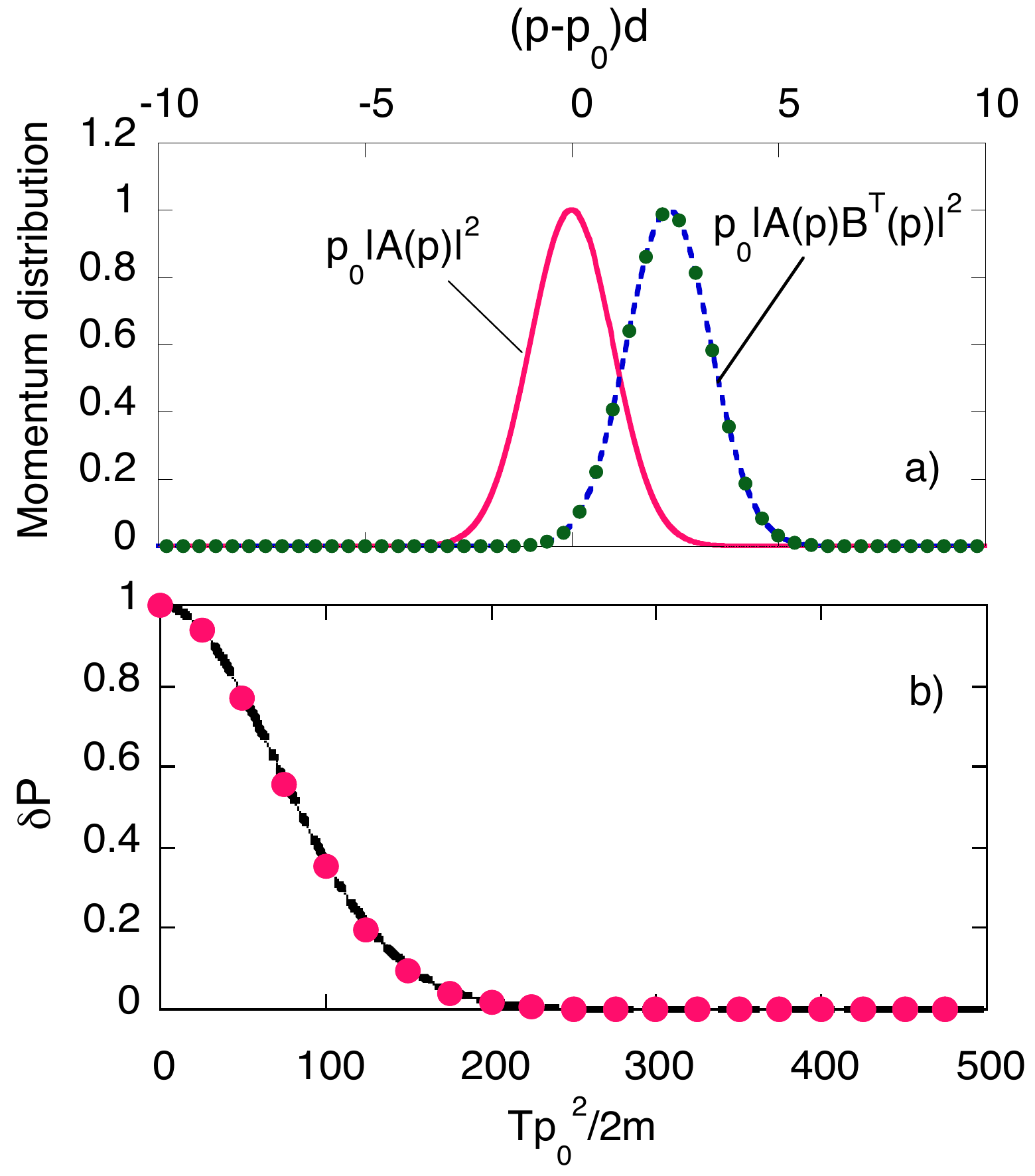}}
\caption{Tunnelling across a broad rectangular barrier with the parameters $p_0d=400$, $p_0^2/2mW=0.5$ and $\sigma=d/2$. (a) the momentum distributions of the incident (solid) and the tunnelled (dashed) wave packets. Both curves are  normalised to unit heights. Also shown by filled circles 
is the approximation (\ref{28}). (b) $\delta P(T)$ for the incidents wave packets (solid), and $\delta P/(P^2T)^2$ for the transmitted ones (filled circles).  }
\label{fig:4}
\end{figure}
\section{Conclusions and discussion}
In summary, contact interaction between atoms can alter tunnelling statistics by making two atoms more likely to exit the barrier on the opposite sides, while leaving the total transmission probability unchanged.
The effect is most visible when the barrier significantly narrows the momentum distribution of the transmitted wave packet, as happens in resonance tunnelling. There the evidence of two atoms colliding in the barrier is produced provided the time between the impacts does not exceed the lifetime of the resonance, often interpreted as the delay experienced by the transmitted particle \cite{REV}-\cite{REV2}. No change in the statistics is predicted for non-resonance tunnelling across a broad potential barrier. In the same sense, this suggests that there is no appreciable delay in the barrier, and that two particles, well separated initially, never encounter each other in a classically forbidden region. Finally, we note that for free bosons the Pauli principle produces the opposite 'bunching' effect, whereby two bosons are more likely to exit the scatterer on the same side.
\section{Acknowledgements:}
DS acknowledges support of the Basque Government (Grant No. IT-472-10), and the Ministry of Science and Innovation of Spain (Grant No. FIS2009-12773-C02-01). LB acknowledges the Russian Fund of Fundamental Investigations (Grant 12-01-00247) and the Saint Petersburg State University (Grant 11.38.666.2013)
 
\end{document}